\documentclass[12pt]{iopart}
\usepackage{bm}
\usepackage{amssymb}
\usepackage{amsfonts}
\usepackage{graphics}
\usepackage{graphicx}
\usepackage{amsthm}
\usepackage{bbm}
\begin{document}

\title[$\;$]{Nonlinear effects on the dynamics of quantum harmonic modes coupled through angular momentum$\!\!\!$}

\author{N. Canosa$^{1}$, R. Rossignoli$^{1,3}$,  Javier Garc\'{\i}a$^{1}$, 
Swapan Mandal$^{1,2}$ and   Kartick Chandra Saha$^{2}$
}

\address{ $^{1}$IFLP-CONICET and Departamento de F\'{\i}sica,  Universidad Nacional de La Plata,
C.C.\ 67, La Plata (1900), Argentina.\\
$^{2}$Department of Physics, Visva-Bharati, Santiniketan-731235, India\\
$^{3}$ Comisi\'on de Investigaciones Cient\'{\i}ficas (CIC), La Plata (1900), Argentina}

\begin{abstract}
We investigate nonlinear effects on  the dynamics of entanglement and 
other  quantum observables in a system of two  harmonic modes  coupled through angular momentum.  The nonlinearity arises from a Kerr-type  anharmonic term in each mode.  The emergence and evolution of entanglement, non-gaussianity, photon number,  photon antibunching and squeezing   are examined  for different initial  coherent product states and  couplings, through exact diagonalization in a truncated basis. 
It is shown that the anharmonic terms, even if weak, can lead to very significant  effects for such initial states,  considerably enhancing and stabilizing entanglement and leading to a non negligible non-gaussianity of the evolved states. They also affect other observables, 
stabilizing the dynamics after an initial transient regime, for not too small initial average populations of each mode.  
Analytic short-time approximate expressions are also  provided. 
\end{abstract}
\maketitle

\section{Introduction}

Nonlinear effects can lead to the development of non trivial phenomena in many  distinct scenarios. In the field of quantum optics 
nonlinear processes  give rise, for instance,  to  spontaneous parametric down conversion \cite{Pe.91},  a fundamental tool for generating entangled photons and hence  of most importance in the field of quantum optics \cite{Mi.08} and quantum information \cite{NC.00,HR.06}. Another well known process associated with nonlinearity is the  Kerr effect \cite{Pe.91,Mi.08,Ta.03}, which plays a fundamental role in the emergence of a wide variety of nonclassical phenomena.  
Kerr nonlinearities were employed  for generating squeezing   in optical fields \cite{Ta.91,Ge.94,Bu.92,Ta.03} and for obtaining    macroscopic  coherent states superpositions (``Schr\"odinger cat states'')   \cite{YS.86,G.99,J.04,K.13}, entangled coherent states  \cite{SR.99, Van.03} and Bell-type  states \cite{Ko.11}. They have also been used for implementing quantum gates for quantum computation  \cite{NC.00,Mil.89,Mun.05,Lin.09,BC.16,Com.18}, quantum teleportation \cite{Vi.00} and other quantum information protocols \cite{Bloo.05,LG.16}, through optical platforms. 
 The effects of Kerr terms  have been recently  investigated in connection with the enhancement of entanglement and  other non classical properties in short chains of non linear oscillators \cite{Ka.16,O.15} as well as in the context of Bose Einstein condensates \cite{He.12,PLK.13,DaNa.17,SO.18}  and  parity-time (${\cal PT}$)  symmetric  systems \cite{Pe.19}. Kerr-like nonlinearities can now be also realized through Rydberg excitations in ultra-cold atomic ensembles (Rydberg nonlinear quantum optics) \cite{Fir.16,Bie.16,Rac.19} and through Josephson junctions in microwave photonics  \cite{Bou.12,Gu.17}.

Motivated by these developments our aim is to investigate, in a system of two harmonic modes interacting through a general quadratic (in the field operators) coupling, the effects of a quartic nonlinearity in each mode. We will focus on the system dynamics, and in particular on the generation and evolution of entanglement and other quantum observables  when starting from an initial  coherent product state.
In an optical system the quadratic coupling considered can account for the exchange of photons between both modes as well as for photon pair creation and annihilation, while the quartic nonlinearity corresponds to a single-mode Kerr effect \cite{Pe.91,HR.06,Ge.94,Bu.92,K.13,Ko.11,Ka.16,Pe.19}. The quadratic interaction between the modes can also represent an angular momentum coupling when the two-mode system describes the motion of a particle in a rotating anisotropic harmonic trap \cite{V.56,RS.80,Blai.86,Li.01,OK.04,Fe.07,RK.09}, or  of a charged particle in a uniform magnetic field within a fixed anisotropic quadratic potential \cite{Fel.70,Ma.94,RK.09}. 
In such scenario the addition of quartic terms represent anharmonic effects in the trapping potential, which can alter the critical properties of the quadratic system, avoiding instabilities \cite{Fe.01,Fe.05}. 
The model considered, suitable for  simulation by optical means \cite{Pe.94}, is then ubiquitous in different scenarios, including  
rapidly rotating Bose-Einstein condensates within the set of lowest Landau level states  \cite{Li.01,OK.04,Fe.07,Fe.01,Fe.05,BD.08,Fe.09,Ho.01,AB.05,AB.09}, quantum dots \cite{Ma.94} and  rotating nuclei \cite{V.56,RS.80,Blai.86}. 
In Bose-Einstein condensates the
addition of a weak quartic confinement potential to the magnetic harmonic trap can be achieved through a suitable detuned laser beam \cite{Stock.04}. 

In previous works we have investigated in detail the present system without the Kerr terms, examining its analytical solution in both stable and unstable sectors and its dynamical phase diagram \cite{RK.09,RR.11,RCR.14,CMR.15}, as well as other aspects such as the generation and control of entanglement between the two modes and squeezing \cite{RCR.14,CMR.15}.  The study  of entanglement and squeezing  arising between two coupled harmonic \cite{R2JMM.11}  and  anharmonic modes  \cite{R2chu.10,R2CC.07,WB.16}, and between two coupled harmonic modes  in an  open regime \cite{R2JO.14,R2GF.16} have  attracted much attention.

Here we will analyze to what extent  the presence of nonlinear terms  affects the evolution of relevant  quantum properties such as the entanglement between the two modes,  squeezing and photon antibunching. We will show that for initial product coherent states the evolution of entanglement and  other observables, which  in the absence of nonlinear effects show a characteristic  oscillatory evolution, experience a considerable change,  evolving  into  an approximately  stable regime after an initial oscillatory transient for not too small values of the initial intensity, i.e. the initial average occupation of each mode. The entanglement generated between the two modes becomes in fact strongly enhanced. We will also examine  the  emergence of non-gaussianity induced by    
such nonlinear terms, showing that it becomes also quite significant.  
 It is  seen as well that  while subpoissonian single mode statistics (antibunching) is present  for  small values of the initial populations,  for larger values  the system evolves towards  superpoissonian statistics after an initial transient. Similarly, squeezing effects  are   noticeable   only for small values of the initial population. We finally  remark that present results are obtained  through an exact diagonalization in a truncated basis, instead of approximate perturbative-based treatments frequently used in similar models \cite{SM.05,Tha.14,Tha.16, AMO.15, Sen.18}, which are accurate just for sufficiently short times. Exact analytic second order short time expressions for the field operators are nevertheless also derived for indicating the initial trend.
 
  \section{The Hamiltonian}
We will consider a system of two harmonic modes with a general quadratic coupling between them, supplemented with a quartic nonlinearity in each oscillator. The Hamiltonian reads 
  \begin{eqnarray}
 H&=&\hbar\omega_{1}\left(a_{1}^{\dagger}a_{1}+\frac{1}{2}\right)+\hbar\omega_{2}\left(a_{2}^{\dagger}a_{2}+\frac{1}{2}\right)\label{H10}\\
&&-\imath\hbar\lambda_{1}\left(a_{2}^{\dagger}a_{1} -a_{1}^{\dagger}a_{2}\right)-\imath\hbar\lambda_{2}\left(a_{1}a_{2}-a_{1}^{\dagger}a_{2}^{\dagger}\right)\label{H11}\\
\hspace*{-0.5cm} &&+\hbar\beta_{1}a_{1}^{\dagger2}a_{1}^{2}+\hbar\beta_{2}a_{2}^{\dagger2}a_{2}^{2}\, ,
\label{H12}
        \end{eqnarray}
    where $\omega_i>0$ denotes the frequency of mode $i$  and $a_i$, $a_i^\dag$ are the corresponding  dimensionless  annihilation
and creation operators 
 obeying the  commutation relations  
$\,\,\left[a_{i},a_{j}^{\dagger}\right]=\delta_{ij},\,\, [a_i,a_j]=[a^\dag_i,a^\dag_j]=0\,$.
The first two  rows  (\ref{H10})--(\ref{H11}) correspond to the quadratic part $H_q$, which includes the quadratic coupling. 
Here $\lambda_1$ is the strength associated with the interchange of bosons and $\lambda_2$ that with pair creation and annihilation. The third  row (\ref{H12})  contains  the Kerr  anharmonicities $\hbar \beta_i (n_i^2-n_i)$ in  each mode, with $\beta_i>0$  the 
corresponding strengths. 
For $\lambda_1=\lambda_2=0$, the system reduces to two independent anharmonic quartic oscillators with total energies  
\begin{equation} 
\begin{array}{rcl}E^0_{n_1,n_2}&=&E^0_{n_1}+E^0_{n_2}\,,\,\hspace{2cm} (\lambda_1=\lambda_2=0)\\
E^0_{n_i}&=&\hbar\omega_i(n_i+\frac{1}{2})+\hbar\beta_i n_i(n_i-1)\,,\;i=1,2\end{array}
\end{equation}
becoming each mode equivalent to a Kerr-like oscillator \cite{OS.19,Mil.86,Ge.94}. The anharmonic effect at no coupling is then to introduce an increasing spacing between the original harmonic levels:  $E^0_{n_i+1}-E^0_{n_i}=\hbar\omega_i+2\hbar\beta_i n_i$, $E^0_{0}=\hbar\omega_i/2$. 

On the other hand, 
the quadratic part (\ref{H10})--(\ref{H11}) can be diagonalized by a suitable bosonic Bogoliubov transformation  in the dynamically stable regions \cite{RK.09,RCR.14,RK.05}. Without loss of generality we can always set 
$\lambda_i>0$ for $i=1,2$ as their signs can be changed by simple local phase transformations ($\lambda_i\rightarrow -\lambda_i$ for $i=1,2$ if $a_1\rightarrow -a_1$,  while $\lambda_2\rightarrow -\lambda_2$ if  $a_i\rightarrow \imath a_i$ for $i=1,2$). Moreover, 
for $a_1\rightarrow \imath a_1$, 
the coupling  adopts the real form  $\hbar\lambda_1(a^\dag_2a_1+a^\dag_1 a_2)+
\hbar\lambda_2(a_1 a_2+a^\dag_1 a^\dag_2)$. 

We use the complex form because a possible realization of the full quadratic part (\ref{H10})-(\ref{H11})  is a system of two harmonic oscillators coupled through an angular momentum term \cite{RK.09,RCR.14}, i.e., 
 \begin{equation}
H_q=\frac{P_{1}^{2}}{2m}+\frac{m\omega_{1}^{2}}{2}Q_{1}^{2}+\frac{P_{2}^{2}}{2m}+\frac{m\omega_{2}^{2}}{2}Q_{2}^{2}-\omega(Q_{1}P_{2}-P_{1}Q_{2})\label{Hq}\,,
\end{equation}
where 
\begin{equation}
Q_{i}=  \sqrt{\frac{\hbar}{2m\omega_{i}}}(a_{i}+a_{i}^{\dagger})\,,\;\;\
P_{i} =  -\imath\sqrt{\frac{\hbar m\omega_{i}}{2}}(a_{i}-a_{i}^{\dagger})
\label{5}
\end{equation}
are the associated coordinates and momenta satisfying 
$[Q_i,P_j]=\imath\hbar\delta_{ij}$, 
 $[Q_i,Q_j]=[P_i,P_j]=0$. Eq.\ (\ref{Hq}) is identical with the quadratic part  (\ref{H10})--(\ref{H11}) for 
\begin{equation}
\lambda_{1}  =  \frac{\omega}{2}\left(\sqrt{\frac{\omega_{1}}{\omega_{2}}}+ \sqrt{\frac{\omega_{2}}{\omega_{1}}}\right)\,,\;\;\lambda_{2}  =  \frac{\omega}{2}\left(\sqrt{\frac{\omega_{1}}{\omega_{2}}}- \sqrt{\frac{\omega_{2}}{\omega_{1}}}\right)\,,\label{la}
\end{equation}
which satisfy $\lambda_1>\lambda_2>0$  if  $\omega>0$ and $\omega_1>\omega_2>0$.  We will assume these conditions in what follows. Here $\omega$ is the angular momentum coupling strength. It  can be considered as a rotational frequency,  such that (\ref{Hq}) represents the cranked Hamiltonian $H-\omega L_z$ $\,(L_z=Q_1 P_2-P_1 Q_2)$ describing the intrinsic motion of a particle in a rotating  anisotropic harmonic trap, in the plane perpendicular to the rotation axis \cite{RK.09,RS.80}. The Hamiltonian  (\ref{Hq}) can describe as well the motion in the $xy$ plane  of a charged particle in a magnetic field ${\bf{H}}$ along  the $z$ axis within an anisotropic quadratic potential. In this case $\omega=e|{\bf{H}}|/(2mc)$ and $m\omega_i^2=K_i+m\omega^2$, with $K_i$ the trap spring constants \cite{RK.09,Ma.94}. 
  
 Let us mention that in an optical realization of the model, 
 the first term of the  quadratic coupling in (\ref{H11}),  associated with the interchange of bosons between the modes, can be realized  by the action of a beamsplitter, while the second  term, which represents photon pair creation and annihilation,  can be provided by a two mode  parametric process \cite{Pe.91,Mi.08,Pe.94}, with the nonlinearities in (\ref{H12}) provided by a Kerr media in each mode \cite{Pe.91,Pe.94}.  In other realizations present quartic terms stem from anharmonicities in the trap spectrum, which can be induced through the addition of a suitable laser trap, 
 as has been implemented  in rotating 
  Bose-Einstein condensates   \cite{Stock.04}. 

The quadratic Hamiltonian (\ref{Hq}) exhibits a rich dynamical phase diagram \cite{RK.09,RR.11,RCR.14}. 
The dynamically and energetically  stable regime where 
$H_q$ is positive definite takes place for 
   $\lambda_1+\lambda_2<\sqrt{\omega_1\omega_2}$,  
 which is equivalent to $\omega< 
 \omega_2$,  i.e. the rotational frequency should not exceed that  of the weaker harmonic mode 
 (this condition is always fulfilled for the charged particle in a magnetic field within a stable trap, as here $\omega_i^2-\omega^2=K_i/m>0$).  
In this regime, $H_q$ can be rewritten as a sum of two normal harmonic modes, 
\begin{equation}
    H_q={\textstyle\hbar\omega_+(a^\dag_+a_++\frac{1}{2})
+\hbar\omega_-(a^\dag_-a_-+\frac{1}{2})}\;\;\;\;\;(\beta_1=\beta_2=0)\,,\label{Hn}
\end{equation}
having energies $E^q_{n_+,n_-}=\hbar\omega_+(n_++1/2)+\hbar\omega_-(n_-+1/2)$, where the normal frequencies $\omega_\pm$ are given by 
\begin{eqnarray}
    \hspace*{0cm}\omega_{\pm}&=&\sqrt{
    \frac{\omega_1^2+\omega_2^2}{2}+\lambda_1^2-\lambda_2^2
    \pm\Delta}\nonumber\\&=&\sqrt{ \frac{\omega_1^2+\omega_2^2}{2}+\omega^2\pm\Delta}\,,\label{www}\\
    \hspace*{0cm}\Delta&=&{\sqrt{(\frac{\omega_1^2-\omega_2^2}{2})^2+\lambda_1^2(\omega_1+\omega_2)^2-\lambda_2^2(\omega_1-\omega_2)^2}}
    	\nonumber\\&=&\sqrt{(\frac{\omega_1^2-\omega_2^2}{2})^2+2\omega^2(\omega_1^2+\omega_2^2)}\,.
    \label{wpm}
\end{eqnarray}
Here (\ref{www})--(\ref{wpm}) are the values for the couplings (\ref{la}). At the border of stability ($\lambda_1+\lambda_2\rightarrow \sqrt{\omega_1\omega_2}$), 
$\omega_-\rightarrow 0$. The normal boson operators are of the form  $a_{\pm}=\sum_{i=1,2}U_{i\pm}a_{i}+V_{i\pm} a^\dag_{i\pm}$ \cite{RK.09,CMR.15}. 
The quartic coupling becomes more complex when expressed in terms of the normal operators $a_{\pm},a^\dag_{\pm}$. It does  not  commute with the boson number of each normal mode, containing terms that create (and destroy) two and also four normal bosons. We also note that for $\omega_1=\omega_2$, $\omega_{\pm}=\sqrt{(\omega_1\pm\lambda_1)^2-\lambda_2^2}$, which in the realization (\ref{Hq}) ($\lambda_1=\omega$, $\lambda_2=0$ in (\ref{la}) become $\omega_{\pm}=\omega_1\pm\omega$, as (\ref{Hq}) 
 commutes with  $L_z$ when $\omega_1=\omega_2$.

\section{Results}
In this section we  present numerical results for the evolution of some relevant quantum observables when the  system starts in  an initial coherent product state. We will   analyze to what extent  the presence of the nonlinear terms  (\ref{H12}) in  $H$ affects   entanglement generation as well  as   squeezing and photon antibunching.
Other indicators of  nonlinear effects  such as  non-gaussianity  will also be  considered.  

For diagonalizing the full Hamiltonian (\ref{H10})-(\ref{H11})--(\ref{H12}), we employed a numerical procedure  consisting in the exact diagonalization of $H$ in a truncated basis comprising the first $m$ states  in each original mode ($n_i\leq m$), with $m$ sufficiently large in order to achieve convergence of the final observables.  
As a check, we have also applied the same procedure in the normal mode basis, by appropriately transforming the quartic coupling, verifying that identical results (within the working tolerance) are obtained. 
Time evolution of an initial state $|\Psi_0\rangle$ is then simply determined as $|\Psi(t)\rangle=\sum_\nu e^{-iE_\nu t/\hbar}\langle \nu|\Psi_0\rangle|\nu\rangle$, 
with $|\nu\rangle$ the exact eigenstates  $H|\nu\rangle=E_\nu|\nu\rangle$ and the sum restricted to the first $m^2$ eigenstates considered.  Time averages of an observable $O$ can then be evaluated as 
\begin{equation}\langle O\rangle_t=\langle\Psi(t)|O|\Psi(t)\rangle=\sum_{\mu,\nu}e^{i(E_\mu-E_\nu)\,t/\hbar}\langle \mu|O|\nu\rangle\,.\end{equation}

We will consider coherent product initial states  
\begin{equation}
    |\Psi(0)\rangle=|\alpha_1,\alpha_2\rangle=e^{-(|\alpha_1|^2+|\alpha_2|^2)/2} e^{\alpha_1 a^\dag_1+\alpha_2 a^\dag_2}|0\rangle\label{chs}\,,
\end{equation}
satisfying 
\begin{equation}
    a_i|\alpha_1,\alpha_2\rangle=
\alpha_i|\alpha_1,\alpha_2\rangle\,, 
\end{equation}
for $i=1,2$, with average number 
$\langle \Psi(0)|a^\dag_i a_i|\Psi(0)\rangle=|\alpha_i|^2$. 
This initial state has then no entanglement between the modes. 
For $\lambda_1=\lambda_2=0$, obviously no entanglement will be generated as there is no coupling between modes. Nonetheless,  the state will not  remain  coherent for $t>0$ if $\beta_i>0$  due to the quartic terms.  

On the other hand, for $\beta_1=\beta_2=0$ 
the Hamiltonian is quadratic and hence a closed analytic evaluation of the time evolution becomes feasible \cite{CMR.15}. 
The Heisenberg field  operators can in this case be explicitly obtained: 
\begin{equation}a_i(t)=e^{iHt/\hbar}a_i e^{-iHt/\hbar}=\sum_{j=1,2}U_{ij}(t)a_j+V_{ij}(t)a^\dag_j\label{eq1}\;\;\;\;\;(\beta_1=\beta_2=0)\, ,
    \end{equation}
with $a^\dag_i(t)=e^{iHt/\hbar}a^\dag_i e^{-iHt/\hbar}=\sum_{j}U_{ij}^*(t)a^\dag_j+V_{ij}^*(t)a_j$, 
where $a_i\equiv a_i(0)$ and the $2\times 2$ matrices $U(t)$ and $V(t)$  are  given in \cite{CMR.15}.  Eq.\ (\ref{eq1}) constitutes  a proper Bogoliubov transformation (such that $[a_i(t),a_j^\dag(t)]=\delta_{ij}$, 
$[a_i(t),a_j(t)]=[a_i^\dag(t),a_j^\dag(t)]=0$ $\forall$ $t$).  Averages at time $t$ of any observable $O$ can then be determined by replacing the operators $a_i, a^\dag_i$ by $a_i(t)$ and $a^\dag_i(t)$ respectively and evaluating the ensuing expression in the initial state (\ref{chs}).  
We have also checked that the numerical procedure employed for the complete Hamiltonian leads in the quadratic case to the same results obtained from the analytic expressions within the working tolerance.

\subsection{Entanglement and non-gaussianity}
We will first analyze the emergence and evolution of entanglement 
between the two modes. It  can be quantified through the entanglement entropy, 
which is the entropy of the reduced state of a single mode: 
 \begin{eqnarray} E_{12}(t)&=&S(\rho_{1}(t))=S(\rho_2(t))\,,\label{E12}
 \end{eqnarray}
where  $S(\rho_i(t))=-{\rm Tr}\,\rho_i(t)\log_2\rho_i(t)$ is the von Neumann entropy and $\rho_{1(2)}(t)={\rm Tr}_{2(1)}\,|\Psi(t)\rangle\langle\Psi(t)|$  are the isospectral  reduced density matrices of each mode. 

In the quadratic case 
$\beta_1=\beta_2=0$, the global state $|\Psi(t)\rangle$ will remain gaussian at all times, implying gaussian single mode reduced densities. The entanglement between the two modes will then be determined solely by the single mode covariance matrix, implying  
 that it will be independent from the values of $\alpha_1,\alpha_2$ determining the initial coherent state, coinciding with that generated from the initial vacuum. 
Explicitly, in the quadratic case Eq.\ (\ref{E12}) becomes 
\begin{eqnarray}S(\rho_i(t))&=&S_g(f_i(t))\,,\;\;\;(\beta_1=\beta_2=0)\label{Eg0}\\
S_g(f_i(t))&=&-f_i(t)\log_2 f_i(t)+(1+f_i(t))\log_2(1+f_i(t))\,,\label{Eg}
 \end{eqnarray}
 where $f_i(t)=\sqrt{(\langle a^\dag_i(t)a_i(t)\rangle-|\langle a_i(t)\rangle|^2+\frac{1}{2})^2-|\langle a_i^2(t)\rangle-\langle a_i(t)\rangle^2|^2}-\frac{1}{2}$,  is the  symplectic eigenvalue of the single mode covariance matrix, with $f_1(t)=f_2(t)$ in the quadratic case. 
 
 We remark, nevertheless,  that in the presence of Kerr terms ($\beta_1>0, \beta_2>0$) Eq.\ (\ref{Eg0}) no longer holds and the generated entanglement is to be computed through Eq.\ (\ref{E12}). It will strongly depend on the initial values of $\alpha_1,\alpha_2$.  Moreover, the difference between (\ref{Eg}) and (\ref{E12}), 
 \begin{equation}
 \Delta S_i(t)=S_g(f_i(t))-S(\rho_i(t))\,,\label{Sgi}
 \end{equation}
 is   an indicator of non-gaussianity of the evolved state. 

   \begin{figure}[ht]
\hspace*{1.25cm}\scalebox{.25}{\includegraphics
{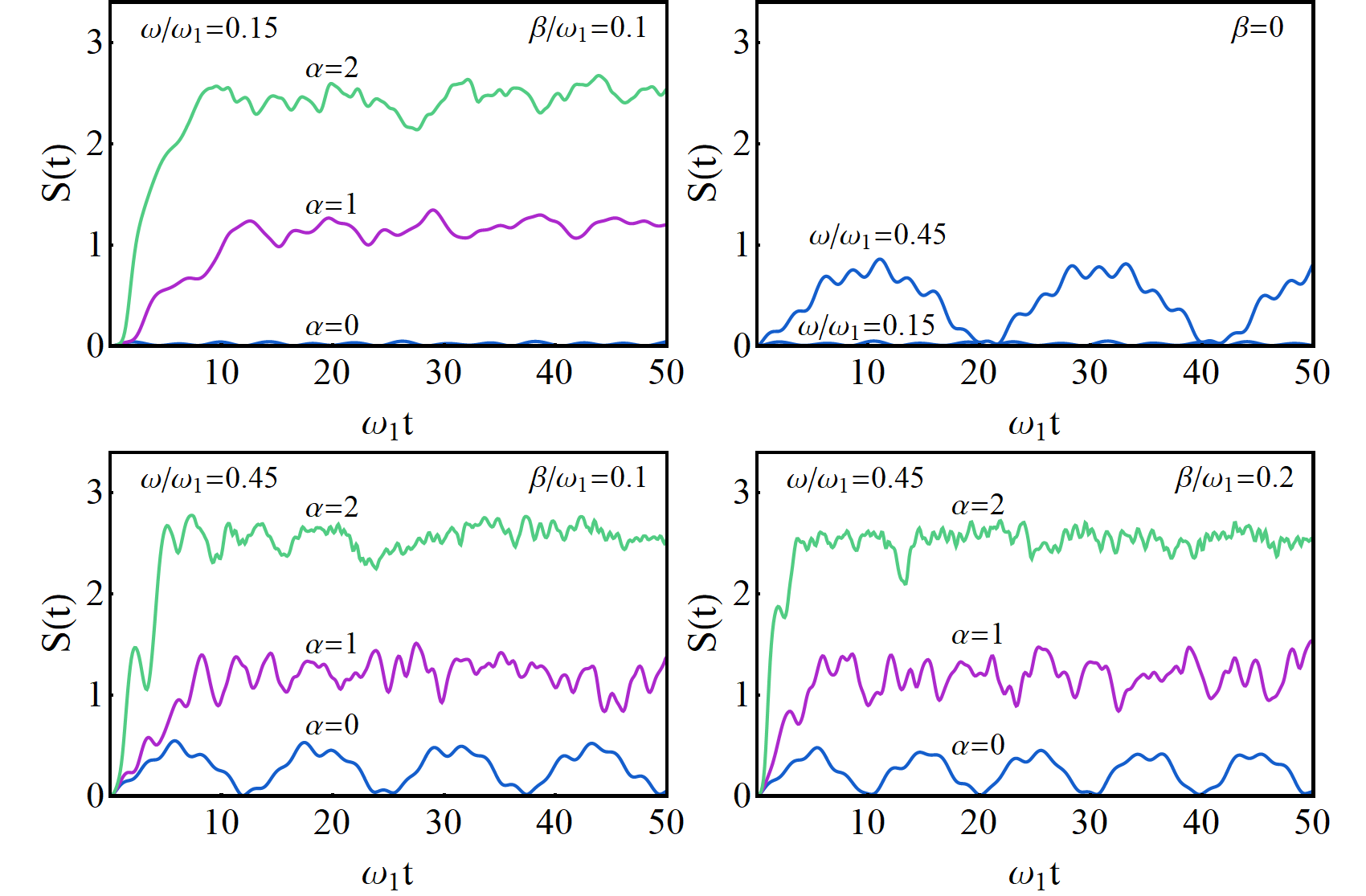}}
\caption{Evolution of the entanglement entropy $S(t)$  
for coherent product initial states $|\alpha,\alpha\rangle$, Eq.\ (\ref{chs}),  and different values of the quartic anharmonic coupling $\beta=\beta_1=\beta_2$ in $H$ for 
$\omega_2=\omega_1/2$ and two values of the quadratic coupling $\omega$ in (\ref{la}).  
For $\beta=0$ (top right panel), entanglement is independent of $\alpha$ and significant just for  sufficiently large $\omega$. In contrast, for $\beta\neq 0$ (top left and bottom panels) entanglement depends strongly on the initial state, stabilizing around  an average value which depends only weakly on $\beta$ and $\omega$ (bottom panels).}
\label{f1}
\end{figure}

Results for the evolution of the entanglement entropy are shown in figure\ \ref{f1}, for  different initial coherent states. We have set $\omega_2=\omega_1/2$  and used two values of  the   coupling $\omega$ in (\ref{H11}): 
 $\omega =0.15\,\omega_1$ (weak quadratic coupling) and  $\omega=0.45\,\omega_1$ (strong quadratic coupling regime, where $\omega$ is close to $\omega_2$ i.e.\  to the instability border of the quadratic case $\beta_i=0$). We have also used  two  different values of the quartic anharmonic coupling,  setting $\beta_1=\beta_2=\beta$. 
 
 It is seen that the presence of quartic terms in  $H$   has a very significant  effect on the evolved entanglement, even for small $\beta$. In the first place 
 the generated entanglement depends strongly on the initial value of $\alpha$, i.e., on the initial average boson number, as  seen in the left top and bottom panels, increasing substantially with $\alpha$. This is in sharp contrast with the quadratic case $\beta_i=0$ (top right panel) where it is independent of $\alpha$, i.e., the same as that obtained when the initial state is the vacuum (an analytical result verified  in the numerical calculations). In the pure quadratic case entanglement from the initial vacuum is generated by the pair creation terms in (\ref{H11}) ($\lambda_2$ coupling), rather than  the $\lambda_1$ coupling,  remaining then small in the weak coupling regime. However, for $\beta\neq 0$ the anharmonic terms become important for increasing $\alpha$ (i.e., nonzero initial $\langle N_i\rangle$, with $N_i=a_i^\dag a_i$),  deviating the evolved state from the gaussian regime and allowing the stronger $\lambda_1$ coupling to play a relevant role, thus increasing the entanglement. 

 It can be also seen in the top left and bottom panels that in the presence of anharmonic terms and $\alpha>0$, after a rapid initial increase entanglement   stabilizes around an average value which at this stage depends only weakly on $\omega$ and $\beta$, as verified in the bottom panels. Only for $\alpha=0$ (vacuum initial state) do the results of the anharmonic case resemble those of the pure quadratic regime, remaining small and oscillating. 

\begin{figure}[ht]
\hspace*{0.25cm}\scalebox{.285}{\includegraphics
{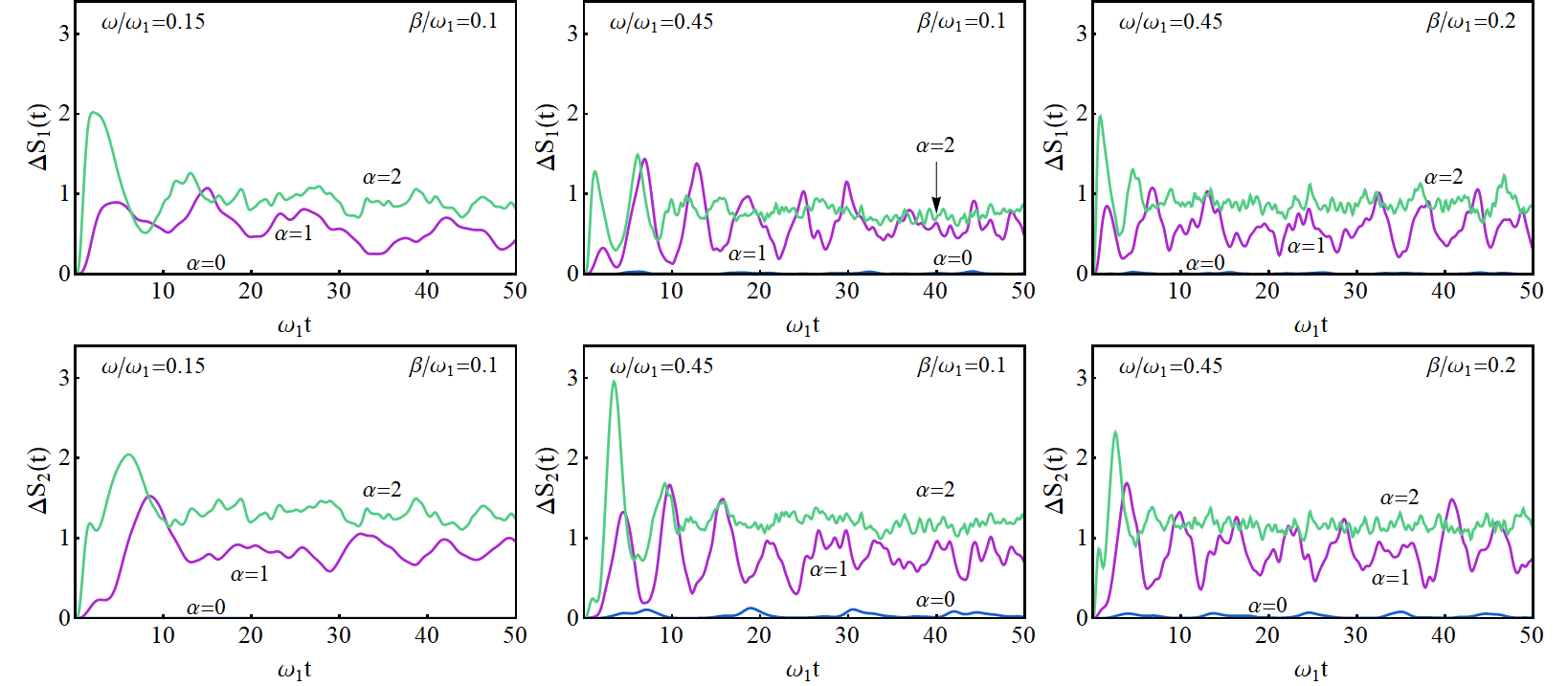}}\caption{The evolution of the non-gaussianity indicator $\Delta S_i(t)$, Eq.\ (\ref{Sgi}), for the first (top row) and  second (bottom row) harmonic mode, for
$\beta/\omega_1=0.1$ (left and middle columns) and $\beta=0.2$ (right column), and  $\omega/\omega_1=0.15$ (left column) and $0.45$ (middle and right columns), and different values of $\alpha$ in the initial coherent state. $\Delta S_i(t)=0$ $\forall$ $t$ for $\beta=0$, remaining small for $\alpha=0$.}
\label{f2}
\end{figure}

Figure \ref{f2} depicts the evolution of $\Delta S_i$, the  indicator of non-gaussianity of the evolved state given in (\ref{Sgi}), for each mode (top  and bottom rows) in the cases of figure\ \ref{f1}.  We compare  different values of $\alpha$ in the initial coherent state both in the weak ($\omega =0.15\,\omega_1$, left column) and strong ($\omega =0.45\,\omega_1$,  middle and right columns) coupling regime for  $\beta=0.1$ (left and middle columns) and $\beta=0.2$ (right column). 
It is seen that $\Delta S_i$ is large for $\alpha>0$ in both modes (slightly larger in the shallower second mode),  
which explains the strong deviation 
 from the gaussian regime of the generated entanglement. Again, after 
 a sharp initial increase, which becomes more prominent as $\alpha$ increases, it tends to stabilize, exhibiting  oscillations which  become less significant as $\alpha$ increases.  
 On the other hand, for  $\alpha=0$, $\Delta S_i$  remains very small, 
 becoming non-negligible only for sufficiently large $\omega$.  This effect is more appreciable in the second mode.

\subsection{Average occupations and squeezing}

\begin{figure}[ht]
\hspace*{1cm}\scalebox{.25}{\includegraphics
{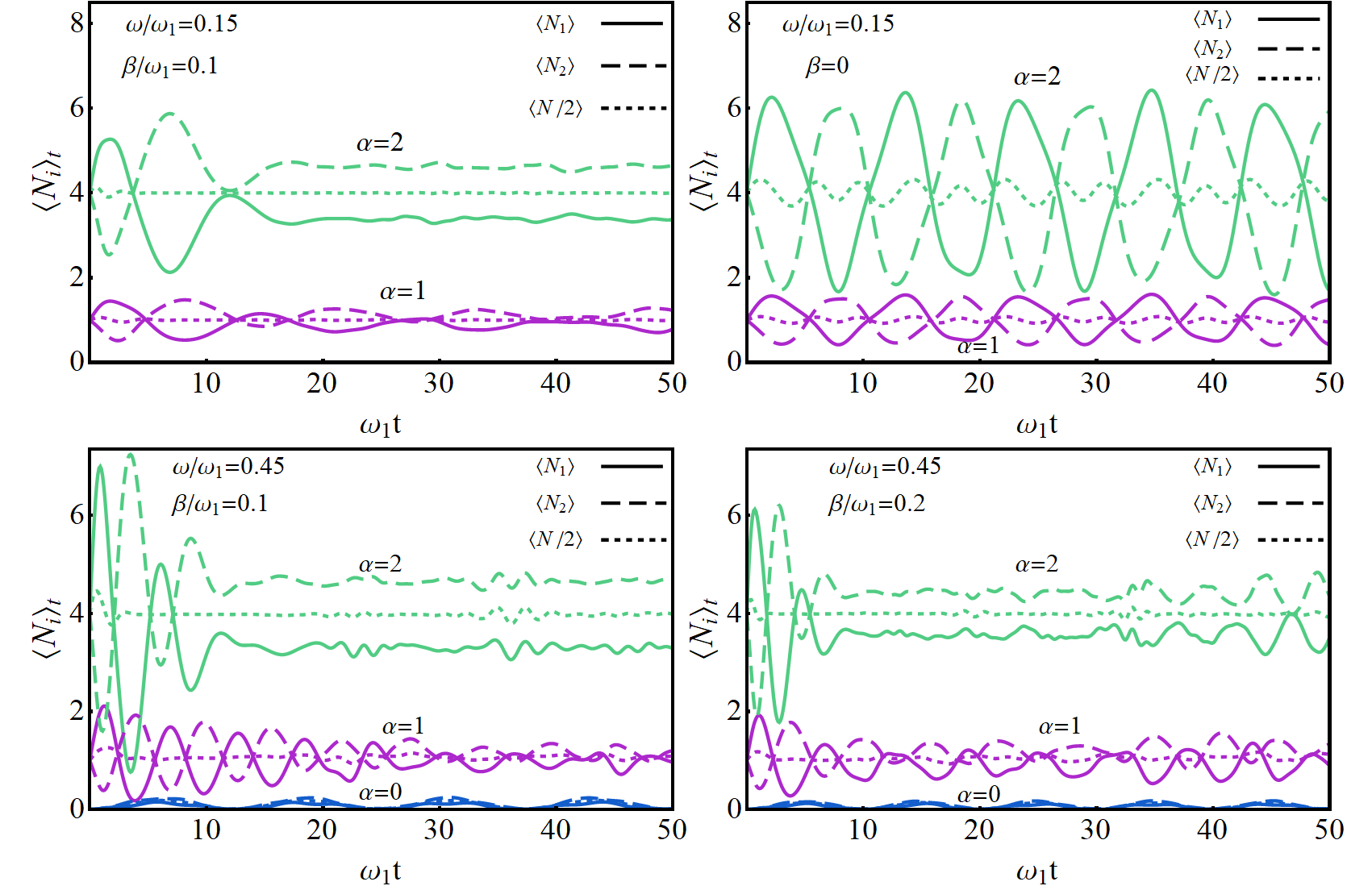}}\caption{The evolution of the average occupation number of each mode $\langle N_i\rangle_t$ (solid and dashed lines) together with their average $\langle N/2\rangle_t$,  with $N=N_1+N_2$ (dotted lines),  
for $\omega/\omega_1=0.15$ (top) and $0.45$ (bottom) and  different values of the anharmonicity $\beta$ and coherent initial state parameter $\alpha$.}
\label{f3}
\end{figure}

 Figure \ref{f3} depicts the evolution of the average occupation number  of  each mode $\langle N_i\rangle$ together with the total average occupation  $\langle N/2\rangle=\langle \frac{N_1+N_2}{2}\rangle$, 
for the cases of figure\ \ref{f1}. In the absence of Kerr terms ($\beta=0$, top right panel) they exhibit a typical oscillatory behaviour reflecting the hopping of bosons between the two modes,  whose amplitude increases with $\alpha$. Moreover, the total average number also exhibits smaller but non-negligible oscillations around the initial value due to the $\lambda_2$ coupling (pair creation and annihilation terms). 

The behaviour changes, however, significantly in the presence of Kerr terms ($\beta>0$) for $\alpha>0$, as seen in the top left and bottom panels.  They lead to an 
 attenuation in the amplitude of the population oscillations 
 of  each mode for $\alpha>0$,
  evolving  from an oscillatory to an approximately  stable regime, particularly as $\alpha$ increases (although  revivals of oscillations may occur for not  too large times as $\omega$ or $\beta$ increases, as seen in the bottom panels). For not too small $\alpha$, after a   short initial oscillatory interval each $\langle N_i\rangle$  stabilizes around  an average value, which lies above the common initial value in the case of  the shallower second mode, and hence below the initial value  in the case of the  steeper first mode, implying a final population transfer. In contrast, the total average occupation remains practically constant (and equal to the initial value) after a very short and limited initial oscillatory interval. Hence, anharmonic effects for $\alpha\gtrsim 1$ essentially downsize those arising from the $\lambda_2$ coupling.   This behaviour 
is similar in  the weak and  strong coupling cases 
(top and bottom left panels), 
although the frequency of the initial oscillations is obviously higher in the strong coupling case. On the other hand, for $\alpha=0$ anharmonic effects remain small, and number oscillations are only visible for strong coupling (bottom panels). 

We also notice that 
\begin{eqnarray}\frac{d\langle N_{i}\rangle}{dt}&=&\frac{\imath}{\hbar}\langle [H,N_i]\rangle=2 {\rm Re}\left[(-1)^{i+1}\lambda_1 \langle a^\dag_2 a_1\rangle 
+\lambda_2\langle a_1 a_2\rangle\right]\,,\label{nit}\end{eqnarray}
for $i=1,2$, which implies $\frac{d\langle N/2\rangle}{dt}=2 \lambda_2{\rm Re}(\langle a_1 a_2\rangle)$  for the average total number. Hence the almost constant behaviour of $\langle N\rangle$ for $\beta>0$ and $\alpha>0$ (for not too small $t$) entails that quartic effects lead essentially to a suppression of ${\rm Re}(\langle a_1 a_2\rangle)$ when the initial population of each mode is non-vanishing, implying opposite variation rates of $\langle N_1\rangle$ and $\langle N_2\rangle$, as verified in top left and bottom panels of figure\ \ref{f3}. Moreover, quartic terms also lead as $t$ increases to a suppression of ${\rm Re}\langle a^\dag_2 a_1\rangle$ for sufficiently large $\alpha$ and small $\omega$, leading to the almost constant stable regime of $\langle N_i\rangle$ seen in top left panel. 

At $t=0$ we obtain from (\ref{nit}), assuming all $\alpha_i$ real,  the initial variation rate 
\begin{equation}\left.\frac{d\langle N_{^1_2}\rangle}{dt}\right|_{t=0}
=2\alpha_1\alpha_2(\pm\lambda_1+\lambda_2)=\pm\alpha_1\alpha_2\omega\left(\frac{\omega_1}{\omega_2}\right)^{\pm 1/2}, 
\label{N12}
\end{equation} 
independent of $\beta$, implying $\left.\frac{d\langle N/2\rangle}{dt}\right|_{t=0}=2\lambda_2 \alpha_1\alpha_2$. 
These results explain the initial trends of these quantities  in figure\ \ref{f3}, i.e.\ 
the initial increase (decrease) of the average occupation of the first (second) mode and the  weaker initial increase of the total average occupation. 

\begin{figure}[ht]
\hspace*{1cm}\scalebox{0.25}{\includegraphics
{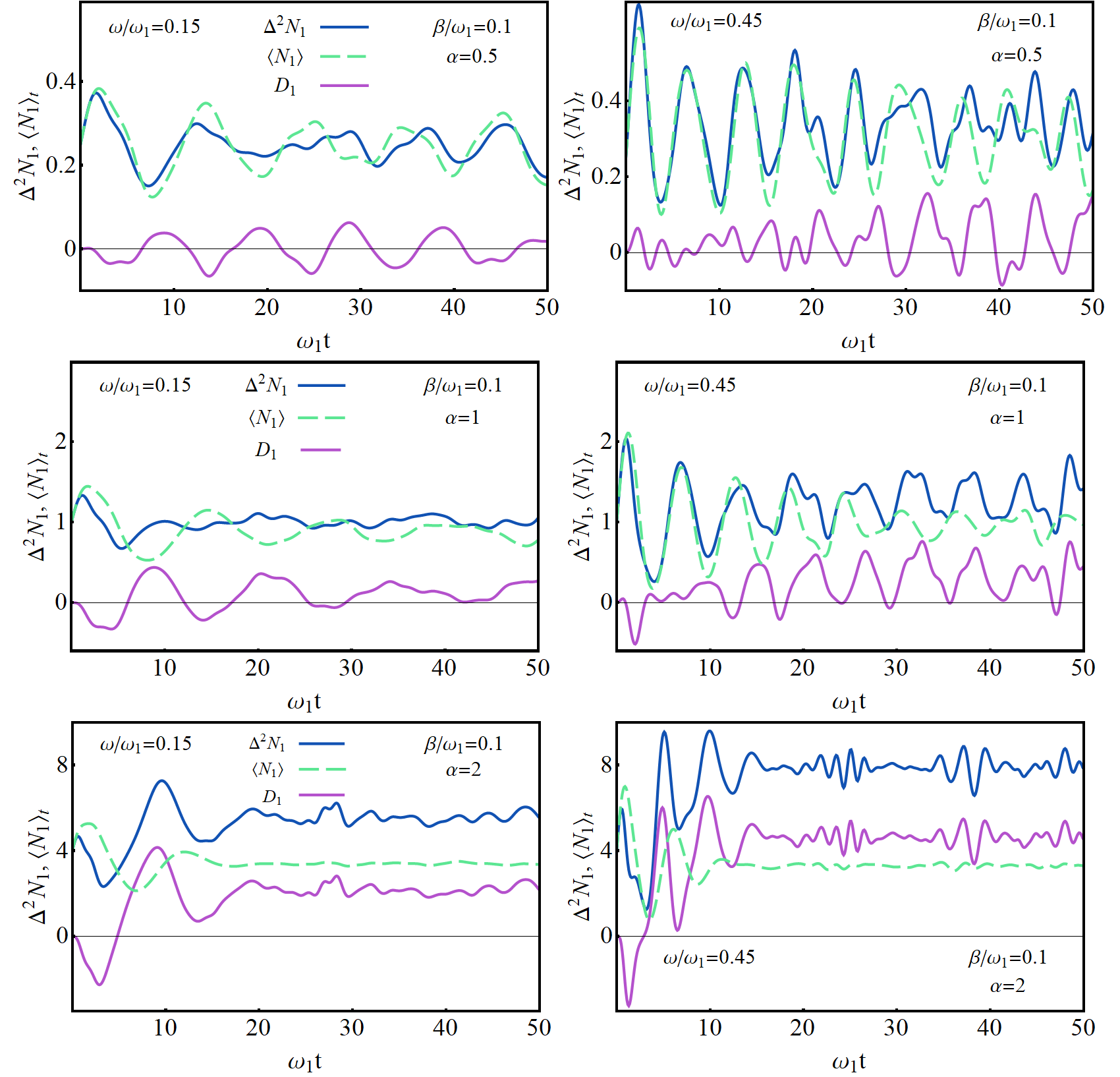}}\caption{Evolution of the occupation number variance  $\Delta^2 N_1=\langle N_i^2\rangle_t-\langle N_i\rangle_t^2$  of the first mode together with  $\langle  N_1\rangle_t$ and the difference 
 $D_1=\Delta^2 N_1-\langle N_1\rangle_t$, Eq.\ (\ref{Di}), for
$\beta/\omega=0.1$  and   $\omega/\omega_1=0.15$ (left panels)  and $0.45$ (right panels), for  an initial coherent state with  $\alpha=0.5$ (top), 
 $1$ (center) and $2$ (bottom).}
\label{f4}
\end{figure}

We also analyze in figure\ \ref{f4} the boson number variance  $\Delta^2 N_i=\langle N_i^2\rangle_t-\langle N_i\rangle_t^2$ of each mode and its difference with the average number, 
\begin{equation}D_i=\Delta^2 N_i-\langle N_i\rangle_t=\langle N_i^2\rangle_t-\langle N_i\rangle_t^2-\langle N_i\rangle_t\label{Di}\,,\end{equation}
  for the first mode (results for the second mode are similar) and two values of the coupling $\omega$. The difference (\ref{Di}) vanishes for coherent states, with $D_i<0$ an indicator of sub-Poissonian single mode photon statistics (antibunching) \cite{Mandel.79,Mira.10} and hence of the quantumness of the state.  This difference is directly related with the Q-Mandel parameter 
  $Q_i=D_i/\langle N_i\rangle_t$, and with the Fano factor $F_i=\Delta^2 N_i/\langle N_i\rangle_t=1+D_i/\langle N_i\rangle_t$, 
  satisfying  $F_i<1$ ($>1$) for  sub-Poissonian (super-Poissonian) statistics. Another related measure of  
  non-classicality is the   second order correlation   function for zero time  delay \cite{Mandel.79,Mira.10} 
$g_i^{(2)}(0)=\langle (a_i^{\dagger})^2 a_i^2\rangle_t/\langle a_i^{\dagger} a_i\rangle_t^2= 1+D_i/\langle N_i\rangle_t^2$, with 
the condition $g_i^{(2)}(0) <1 $ corresponding to  antibunching \cite{KDM.77}. 

 We have here  also included results for  a low but nonzero value of $\alpha$ in the top panels, in order to see the presence and extent of the  antibunching effects.   It is clear that
 antibunching   arises any time $\langle N_1\rangle_t$ exceeds $\Delta^2N_1$. For small $\alpha$, this effect occurs almost periodically for  short time intervals, in part as a result of dephasing between $\Delta^2 N_1$ and $\langle N_1\rangle_t$ as $t$ increases, 
 as seen in the top panels. Higher values of $\omega$ narrow the intervals where the antibunching effect appears (right panels).
  For larger values of $\alpha$ antibunching diminishes as time increases (as seen in the central panel for $\alpha=1$),  becoming restricted to a small initial interval for larger values of $\alpha$, as seen in the bottom panels for $\alpha=2$. Anharmonic effects then lead for present conditions essentially to  superpoissonian single mode statistics ($D_i>0$)  in the  stable phase emerging for larger times. 
  
  It is also seen that for large $\alpha$,  the evolution of $\Delta^2N_1$ and $D_1$  shows a  behaviour similar to that of $\langle N_1\rangle_t$ (figure\ \ref{f3}),  in the sense of exhibiting a short  initial oscillatory regime which rapidly evolves into an approximately stable regime with small fluctuations around an average value, as seen in the bottom panels, which may include a  sequel of small amplitude oscillations around the average  in the strong coupling case (bottom right panel). 
  
  Using Eq.\ (\ref{nit}), we notice that the exact variation rate of the number fluctuation of each mode, and of the quantity (\ref{Di}) is given by 
    \begin{eqnarray}\frac{d\,\Delta^2 N_i}{dt}&=&
    2 {\rm Re}\langle \{N_i-\langle N_i\rangle,(-1)^{i+1}\lambda_1  a^\dag_2 a_1+\lambda_2 a_1 a_2\}\rangle\label{nit2}\,,\\
    \frac{d D_i}{dt}&=&
    2 {\rm Re}\langle \{N_i-\langle N_i\rangle-\frac{1}{2},(-1)^{i+1}\lambda_1  a^\dag_2 a_1+\lambda_2 a_1 a_2\}\rangle\label{nit3}\,,\end{eqnarray}
  where $\{\;,\;\}$ denotes the anticommutator. 
  At $t=0$ (and for real $\alpha_i$) these expressions yield an initial rate 
  $\left.\frac{d\,\Delta^2 N_i}{dt}\right|_{t=0}=2\alpha_2\alpha_1[(-1)^{i+1}\lambda_1+\lambda_2]=\left.\frac{d\langle N_i\rangle}{dt}\right|_{t=0}$, as verified in all panels, and hence 
    $\left.\frac{d D_i}{dt}\right.|_{t=0}=0$.

\begin{figure}[ht]
\hspace*{1cm}\scalebox{0.25}{\includegraphics
{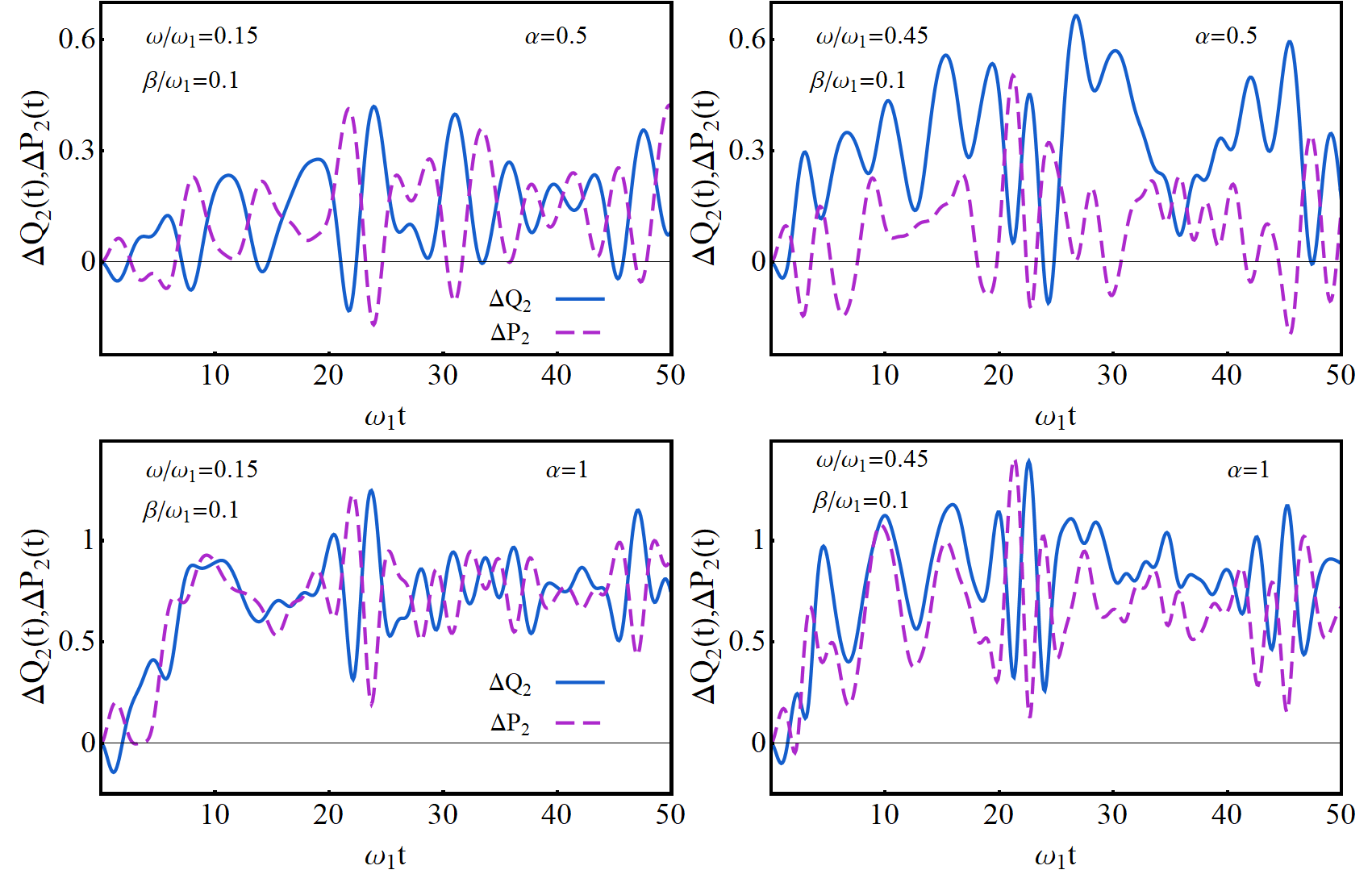}}\caption{Evolution of the shifted squeezing ratios $ \Delta Q_2(t)$ and $ \Delta P_2(t)$ for  
$\beta=0.1$  and $\omega/\omega_1=0.15$ (left panels) and $0.45$ (right panels)  for an  initial coherent state with  $\alpha=0.5$ (top panels) 
and $\alpha=1$ (bottom panels).}
\label{f5}
\end{figure}

  Finally we analyze in figure\ \ref{f5} results for another indicator of single mode quantumness, the shifted squeezing ratios 
  \begin{equation}\Delta O_j= \sqrt{\frac{\langle O_j^2\rangle_t-\langle O_j\rangle_t^2}{\langle O_j^2\rangle_{0}-\langle O_j\rangle_{0}^2}}-1\,,
  \label{R}\end{equation} 
  where  $O_j$ stands for $Q_j$ or $P_j$ and 
  $\langle O_j\rangle_{0}$ denotes the initial expectation value of $O_j$. 
  We show results for  $\Delta Q_i(t)$ and $\Delta P_i(t)$ for the second (shallower) mode, where  the squeezing effects, i.e. negative values of the previous quantities, are more clearly seen. 
  
  The top panels  depict their evolution 
    for small $\alpha=0.5$ and $\beta/\omega_1=0.1$, where squeezing still occurs in short time intervals,   as in the $\beta=0$ case \cite{CMR.15}. 
    In the weak coupling regime $\omega/\omega_1=0.15$ (left panel) both quantities exhibit similar  squeezing effects, although for stronger coupling  $\omega/\omega_1=0.45$ (right panel) they become more appreciable in $P_2$. On the other hand, for higher $\alpha$, i.e.\ higher mode population,  squeezing occurs only in a short initial interval, as seen in the bottom panels. As previously discussed, for large $\alpha$ anharmonic effects become more important and lead for present conditions  
     to a more ``classical-like'' behaviour at the single mode level, with no squeezing. 
     Both shifted ratios fluctuate around a positive average value for not too small times. This behaviour is also  in agreement with the deviation from the gaussian regime   and hence from minimum uncertainty ($(\Delta Q_i+1)(\Delta P_i+1)=1)$ induced by the anharmonic terms.  
     We also mention that the inclusion of present Kerr terms stabilize the dynamics in the region where the quadratic system becomes unstable,  in agreement with the effect of other quartic anharmonicities \cite{Fe.01,Fe.05,Stock.04}.

\subsection{Second order expansion and short time approximation}
Finally we provide results derived from the second order short-time
approximation, which can be used for determining the initial trend. The approach has been employed for obtaining approximate solutions of systems 
of coupled nonlinear differential equations for sufficiently short times \cite{Kodusek,Mandal1,SM.05,perina1,Pathak,perina3}, and used to investigate  quantum statistical properties of the input
radiation field involving the parametric generation \cite{Kodusek}. 
And with similar methods the quantum
entanglement, squeezing and the non-classical distribution of photon-phonon 
in Raman processes were also investigated \cite{Pathak}. It was also found useful for examining the  quantum statistical properties of nonlinear optical couplers \cite{perina3}. 

The exact equations of motion for the Heisenberg field operators
$a_{i}(t)= e^{\imath H t/\hbar}a_{i}(0) e^{-\imath H t/\hbar}$ are
\begin{equation}
\begin{array}{lcl}
\dot{a_1}& =\frac{\imath}{\hbar} [H, a_1(t)] =& -\imath\omega_{1}a_{1}+\lambda_{1}a_{2}+\lambda_{2}a_{2}^{\dagger}-2\imath\beta_{1}a_{1}^{\dagger}a_{1}^{2}\\ 
\dot{a_{2}}&=\frac{\imath}{\hbar} [H, a_2(t)]   = & -\imath\omega_{2}a_{2}-\lambda_{1}a_{1}+\lambda_{2}a_{1}^{\dagger}-2\imath\beta_{2}a_{2}^{\dagger}a_{2}^{2}
\end{array}\label{a8}\,,
\end{equation}
where $a_i\equiv a_i(t)$. These equations are obviously 
nonlinear in the field operators for  $\beta_i>0$. From (\ref{a8}) we can obtain 
the second derivatives as $\ddot{a}_i  = (\frac{\imath}{\hbar})^2[H,[H, a_i]]$:  
\begin{equation}
\begin{array}{lcl}
 \ddot{a}_{1} & = & -(\omega_{1}^{2}+\lambda_{1}^{2}-\lambda_{2}^{2})a_{1}-\imath\lambda_{1}(\omega_{1}+\omega_{2})a_{2}-\imath\lambda_{2}(\omega_{1}-\omega_{2})a_{2}^{\dagger}
 -4\beta_{1}\omega_{1}a_{1}^{\dagger}a_{1}^{2}\\
 &  &- 2\imath\beta_{2}a^\dag_2(\lambda_{1}a_{2}-\lambda_{2}a_{2}^{\dagger})
 a_{2}
 -4\imath\beta_{1}a^\dag_1a_1(\lambda_{1}
 a_{2}+\lambda_{2}a_{2}^{\dagger})  -2\imath\beta_{1}a_1^2(\lambda_{1}a_{2}^{\dagger}+\lambda_2 a_2)\\
 &  &-4\beta_{1}^{2}a_{1}^{\dagger}a_{1}a_{1}^{\dagger}a_{1}^{2}\\
 \ddot{a}_{2} & = & -(\omega_{2}^{2}+\lambda_{1}^{2}-\lambda_{2}^{2})a_{2}+\imath\lambda_{1}(\omega_{1}+\omega_{2})a_{1}+\imath\lambda_{2}(\omega_{1}-\omega_{2})a_{1}^{\dagger}-4\beta_{2}\omega_{2}a_{2}^{\dagger}a_{2}^{2}\\
 &  &+ 2\imath\beta_{1}a^\dag_1(\lambda_{1}a_{1}+\lambda_{2}
 a_{1}^{\dagger})a_1+4\imath\beta_{2}a^\dag_2a_2
 (\lambda_{1}a_{1}-\lambda_2 a_{1}^{\dagger}) 
 +2\imath\beta_{2}a_2^2(\lambda_{1}a_{1}^{\dagger}-\lambda_2 a_1)\\
 &  & -4\beta_{2}^{2}a_{2}^{\dagger}a_{2}a_{2}^{\dagger}a_{2}^{2}
\end{array}\label{a10}\,.
\end{equation}
The first terms of 
the  Taylor series of $a_i(t)$ around $a_{i}(0)$  
are then given by 
\begin{equation}
a_{i}(t)=a_{i}(0)+t\dot{a}_{i}(0)+\frac{t^{2}}{2}\ddot{a}_{i}(0)+O(t^{3})\,.\label{a11}
\end{equation}
and the second order short time approximation is obtained neglecting terms $O(t^3)$. Setting now $a_{i}(0)=a_i$ we  obtain 
\begin{equation}
\begin{array}{rcl}
 a_{1}(t) & = & [1-\imath\omega_{1}t-(\omega_{1}^{2}+\lambda_{1}^{2}-\lambda_{2}^{2})\frac{t^{2}}{2}+\ldots]a_{1}+
[\lambda_{1}t-\imath\lambda_{1}(\omega_{1}+\omega_{2})\frac{t^{2}}{2!}+\ldots]a_{2}\\
 && +  [\lambda_{2}t-\imath\lambda_{2}(\omega_{1}-\omega_{2})\frac{t^{2}}{2!}+\ldots]a_{2}^{\dagger}-[2\imath\beta_{1}t+4\beta_{1}(\beta_{1}+\omega_{1})\frac{t^{2}}{2!}]a_{1}^{\dagger}a_{1}^{2}\\
 & &+[ 2\imath\beta_{2}a^\dag_2(-\lambda_{1}a_{2}+\lambda_{2}
 a_{2}^{\dagger})a_{2}-4\imath\beta_{1}a^\dag_1 a_1(\lambda_{1}a_{2}+\lambda_{2}a_{2}^{\dagger})\\
 & &-2\imath\beta_{1}a_1^2(\lambda_{1}a_{2}^{\dagger}+
 \lambda_{2}a_{2})-4\beta_{1}^{2}a_{1}^{\dagger2}a_{1}^{3}]\frac{t^{2}}{2!}+\ldots\\
 a_{2}(t) & = & [1-\imath\omega_{2}t-(\omega_{2}^{2}+\lambda_{1}^{2}-\lambda_{2}^{2})\frac{t^{2}}{2}+\ldots]a_{2}+[-\lambda_{1}t+\imath\lambda_{1}(\omega_{1}+\omega_{2})\frac{t^{2}}{2!}+\ldots]a_{1}\\
 &  &+ [\lambda_{2}t+\imath\lambda_{2}(\omega_{1}-\omega_{2})\frac{t^{2}}{2!}+\ldots]a_{1}^{\dagger}-[2\imath\beta_{2}t+4\beta_{2}(\beta_{2}+\omega_{2})\frac{t^{2}}{2!}]a_{2}^{\dagger}a_{2}^{2}\\
 && + [2\imath\beta_{1}(\lambda_{1}a_{1}^{\dagger}a_{1}+\lambda_{2}a_{1}^{\dagger})a_1+4\imath\beta_{2}a_2^\dag a_2(\lambda_{1}a_{1}-\lambda_{2}a_{1}^{\dagger})\\
 &  &+ 2\imath\beta_{2}a_2^2(\lambda_{1}a_{1}^{\dagger}-\lambda_{2}a_{1})-4\beta_{2}^{2}a_{2}^{\dagger2}a_{2}^{3}]\frac{t^{2}}{2!}+\ldots
\end{array}\label{12}\,.
\end{equation}
  By taking
the Hermitian conjugate of  (\ref{12}) we obtain the
creation operators for the two field modes. The
commutation relations 
$[a_{i}(t),a_{j}^{\dagger}(t)]=\delta_{ij}$ 
are verified up to second order. 
These  expressions can be used to determine the initial trend of the evolution of any observable. 
 For instance, the population of the first mode  $N_1(t)=a^\dag_1(t)a_1(t)$ is given by 
\begin{equation}
     \begin{array}{rcl}
    N_1(t)    &=&a^\dag_1 a_1+\left[\lambda_1(a^\dag_1 a_2 +a^\dag_2 a_1)+\lambda_2(a_1 a_2+a^\dag_1a^\dag_2)\right]t
    +\left[-2(\lambda_1^2-\lambda_2^2)a^\dag_1 a_1\right.\\
  &&+\imath\lambda_1(\omega_1-\omega_2)(a^\dag_1a_2-a^\dag_2a_1)+\imath\lambda_2(\omega_1+\omega_2)(a^\dag_1a^\dag_2-
  a_1 a_2)+2\lambda_1\lambda_2(a^{\dag 2}_2+a_2^2)\\
  &&-2\imath\{\beta_1 a^\dag_1[\lambda_1 (a_1^2a^\dag_2 -a^{\dag }_1 a_1 a_2)+\lambda_2(a_1^2a_2
  -a^{\dag }_1 a_1 a^\dag_2)]\\
 &&\left.+\beta_2 a^{\dag }_2[\lambda_1 (a_1a^{\dag }_2a_2-
  a^\dag_1  a_2^2)+\lambda_2(a^\dag_1a^{\dag }_2a_2-
  a_1 a_2^2)]\}\right]\frac{t^2}{2}+\ldots
  \end{array}\label{nidt}\,.
\end{equation}
We then see that while absent at first order, 
at second order there is already an interplay between the quartic terms and the quadratic coupling (terms $\propto \beta_i \lambda_j$).  And for an initial product coherent  state, the average population $\langle N_1(t)\rangle$ can be obtained by replacing $a_i$ ($a^\dag_i$) by $\alpha_i$ ($\alpha_i^*$) in (\ref{nidt}). 
In particular, for $\alpha_i$ real,  $\langle N_1(t)\rangle=\alpha_1^2+2\alpha_1\alpha_2(\lambda_1+\lambda_2)t-[(\lambda_1^2-\lambda_2^2)\alpha_1^2-2\lambda_1\lambda_2
\alpha_2^2]t^2+\ldots$, which extends result (\ref{N12}). 

\section{Conclusions}
We have investigated the nonlinear effects of quartic anharmonic terms in the dynamics of entanglement and other quantum observables in a system of two harmonic modes interacting through an angular momentum coupling. The main result is 
that such terms do have a very significant effect on the system dynamics for initial  coherent product states. 
Despite not directly involved in the coupling between both modes, they are able to considerably enhance the generated entanglement between both modes in comparison with the pure quadratic case, as seen in figure\ \ref{f1},   stabilizing it around a nonzero value after a rapid initial increase if the initial average population $|\alpha_i|^2$ of each mode is not too small. Such effect is accompanied by the emergence  of a  non-negligible non-gaussianity, as shown in figure\ \ref{f2}. Moreover, the same effects are seen in the evolution of the average population of each mode (figure\ \ref{f3}),  which rapidly approaches a rather steady regime after an initial  oscillatory transient if $|\alpha_i|^2$ is not too small,  implying a final population transfer between the modes. On the other hand, sub-Poissonian statistics (antibunching) and squeezing in the original modes coordinates become suppressed after a short initial period. These results entail that even small or moderate anharmonicities of the present type in each mode may have deep effects 
in the generated entanglement and in the system dynamics, 
and can be important for an improved control and stability of the system. Approximate analytic treatments for correctly describing these effects beyond short times, and inclusion of environment effects, are currently under investigation. 
\section*{Acknowledgment}
We are thankful to  The World Academy of Sciences
(TWAS), Trieste, Italy and CONICET of Argentina, for
financial support through TWAS-UNESCO fellowship program.
Authors also acknowledge support from CONICET (NC, JG, SM) and
CIC (RR) of Argentina. Work supported  by CONICET PIP Grant 112201501-00732.

\end{document}